# A new lower bound on the independence number of a graph

O.Kettani
Faculté des sciences et techniques-Mohammedia-Maroc

**Abstract**

For a given connected graph G on n vertices and m edges, we prove that its independence number $\alpha(G)$ is at least $((2m+n+2)-((2m+n+2)^2-16n^2)^{½})/8$.

## Intoduction

Let G=(V,E) be a connected graph G on $n=|V|$ vertices and $m=|E|$ edges.
For a subgraph H of G and for a vertex $i\in V(H)$, let $d_H(i)$ be the degree of i in H and let $N_H(i)$ be its neighbourhood in H. Let $\delta(H)$ and $\Delta(H)$ be the minimum degree and the maximum degree of H, respectivly. A subset X of V is called independent if its vertices are mutually non-adjacent. The independence number $\alpha$ (G) is the largest cardinality among all independent sets of G.
The problem of finding an independent set of maximum cardinality is know to be NP –complete[1]. Some approximation algorithms was designed to tackle this problem, among them, the well know MIN algorithm [4], which can be implemented in time linear in n and m :

$G_1$ :=G, j :=1
While $V(G_j)\neq\emptyset$ do
Begin
    Choose $i_j \in V(G_j)$ with $d_{Gj}(i_j)=\delta(G_j)$, delete $\{i_j\}\cup N_{Gj}(i_j)$ to obtain $G_{j+1}$ and set j :=j+1 ;
End ;
k :=j-1
stop.

Let $k_{MIN}$ be the smallest k the algorithm MIN provides for a given connected graph G.
Harant [3] proved that $\alpha(G) \geq k_{MIN} \geq ((2m+n+1)-((2m+n+1)^2-4n^2)^{½})/2$.
The purpose of the present note is to improve this lower bound.

**Claim** : For a given connected graph G on n vertices and m edges,
$$\alpha(G) \geq ((2m+n+2)-((2m+n+2)^2-16n^2)^{½})/8.$$

The proof starts with the inequality (1) proved by Harant [3] :

$$k_{MIN} \geq n^2/\left(2m+n-\sum_{i\in I_n}(d_G(i)-\delta(G_j))\right) \qquad (1)$$

and uses a variation of the one given by Halldorson [2].

For $j=1,\ldots, k_{MIN}$ , let $d_{G_j}(i_j)$ be the degree in the remaining graph of the j–th vertice choosed at the j–th iteration of the algorithm MIN. The number of vertices deleted in the j–th iteration is thus $1+ d_{G_j}(i_j)$ and the sum of the degrees of the $1+ d_{G_j}(i_j)$ vertices deleted is at least $(1+ d_{G_j}(i_j))d_{G_j}(i_j)$. Thus the number of edges removed in the j–th iteration is at least $(1+ d_{G_j}(i_j))d_{G_j}(i_j)/2$.
Let X be an independent set of G of maximum cardinality $\alpha$ , and let $k_j$ be the number of vertices among the $1+ d_{G_j}(i_j)$ vertices deleted in the j–th iteration that are also contained in X.

Then $\sum_{j=1}^{k_{MIN}} k_j = \alpha$

Since X is edgeless, and G is connected then the number of edges removed in the j–th iteration ($j=1,\ldots, k_{MIN} -1$) is at least :

$$\binom{1+ d_{G_j}(i_j)}{2}+\binom{k_j}{2} +1$$

(for $j=1 ,\ldots, k_{MIN} -1$ , there is at least one edge between $N_{G_j}(i_j)$ and $G_{j+1}$, because G is supposed connected).

In the $k_{MIN}$ –th iteration, at least

$$\binom{1+ d_{G_j}(i_{kMIN})}{2}+\binom{k_{kMIN}}{2}$$

edges are removed

Hence we obtain the following inequality :

$$m\geq \sum_{j=1}^{k_{MIN}-1} \left(\binom{1+ d_{G_j}(i_j)}{2}+\binom{k_j}{2} +1\right) +\binom{1+ d_{G_j}(i_{kMIN})}{2}+\binom{k_{kMIN}}{2}$$

then :

$$2m\geq 2k_{MIN} -2+\sum_{j=1}^{k_{MIN}} \left((1+ d_{G_j}(i_j))\, d_{G_j}(i_j) \right)+\sum_{j=1}^{k_{MIN}} k_j +\sum_{j=1}^{k_{MIN}} (k_j)^2$$

consequently :

$$2m \geq 4k_{MIN}-2 + \sum_{j=1}^{k_{MIN}} \left((1+ d_{Gj}(i_j))\, d_{Gj}(i_j)\right) \qquad (2)$$

On the other hand :

Since $\forall\ (j,j') \in \{1, \ldots, k_{MIN}\},\ j \neq j' \Rightarrow \left(\{i_j\} \cup N_{Gj}(i_j)\right) \cap \left(\{i_{j'}\} \cup N_{Gj'}(i_{j'})\right) = \varnothing$ and

$$I_n = \bigcup_{j=1}^{k_{MIN}} \left(\{i_j\} \cup N_{Gj}(i_j)\right) = \{1,\ldots,n\}$$

then

$$\sum_{i \in I_n} \delta(G_j) = \sum_{j=1}^{k_{MIN}} \ \sum_{i \in \{i_j\} \cup N_{Gj}(i_j)} \delta(G_j)$$

$$\sum_{i \in I_n} \delta(G_j) = \sum_{j=1}^{k_{MIN}} (1+ d_{Gj}(i_j))\delta(G_j) \ \leq \ \sum_{j=1}^{k_{MIN}} \left((1+ d_{Gj}(i_j)) d_{Gj}(i_j)\right)$$

thus

$$\sum_{i \in I_n} (d_G(i) - \delta(G_j)) = 2m - \sum_{i \in I_n} \delta(G_j) \geq 2m - \sum_{j=1}^{k_{MIN}} \left((1+ d_{Gj}(i_j)) d_{Gj}(i_j)\right)$$

by using inequality (2) we get :

$$\sum_{i \in I_n} (d_G(i) - \delta(G_j)) \geq 4k_{MIN}-2$$

then inequality (1) implies :

$$k_{MIN} \geq n^2/(2m+n+2-4k_{MIN})$$

and consequently :

$$k_{MIN} \geq \left((2m+n+2) - ((2m+n+2)^2-16n^2)^{1/2}\right)/8.$$

## Conclusion

This note presented an improved lower bound on the independence number of a graph, and as a future work, our intention is to prove that this bound is optimale for an important class of graphs.